\documentclass[11pt,journal, onecolumn, draftcls]{IEEEtran}
\usepackage{amsmath}
\usepackage{epsfig}
\usepackage{amssymb}
\usepackage{graphicx}
\usepackage{hyperref}
\usepackage{color}
\usepackage{subfigure}

\begin{document}
\title{Moving Target Parameters Estimation in Non-Coherent
{MIMO} Radar Systems}
\author{Aboulnasr~Hassanien,~\IEEEmembership{Member,~IEEE},
Sergiy A.~Vorobyov\thanks{S.~A.~Vorobyov is the corresponding
author. A.~Hassanien and S.~A.~Vorobyov are with the Department of
Electrical and Computer Engineering, University of Alberta,
Edmonton, AB, T6G~2V4 Canada; e-mail: \{{\tt hassanie,
sergiy.vorobyov\}@ualberta.ca}. A.~B.~Gershman was with the
Communication Systems Group, Darmstadt University of Technology,
Merckstr. 25, D-64283 Darmstadt, Germany. A.~B.~Gershman passed
away in Aug. 2011.

This work is supported in parts by the Natural Science and
Engineering Research Council (NSERC) of Canada and the European
Research Council (ERC) Advanced Investigator Grants program under
Grant 227477-ROSE.},~\IEEEmembership{Senior Member,~IEEE}, and
Alex~B.~Gershman,~\IEEEmembership{Fellow,~IEEE}}
\vspace{-1cm}
\maketitle
\vspace{-1cm}
\begin{abstract}
The problem of estimating the parameters of a moving target in
multiple-input multiple-output (MIMO) radar is considered and a
new approach for estimating the moving target parameters by making
use of the phase information associated with each transmit-receive
path is introduced. It is required for this technique that
different receive antennas have the same time reference, but no
synchronization of initial phases of the receive antennas is
needed and, therefore, the estimation process is non-coherent. We
model the target motion within a certain processing interval as a
polynomial of general order. The first three coefficients of such
a polynomial correspond to the initial location, velocity, and
acceleration of the target, respectively. A new maximum likelihood
(ML) technique for estimating the target motion coefficients is
developed. It is shown that the considered ML problem can be
interpreted as the classic ``overdetermined'' nonlinear
least-squares problem. The proposed ML estimator requires
multi-dim\-en\-s\-ional search over the unknown polynomial
coefficients. The Cram\'er-Rao Bound (CRB) for the proposed
parameter estimation problem is derived. The performance of the
proposed estimator is validated by simulation results and is shown
to achieve the CRB.
\end{abstract}


\section{Introduction}
The detection and parameter estimation of moving targets is one of
the most important radar applications
\cite{RichardCurry}--\cite{KlemmBook}. The moving target
parameters of interest can be the radar cross section (RCS),
Doppler frequency, range/location, velocity, acceleration, etc.
\cite{Skolnik}. In conventional single antenna radar, the target
RCS and range are measured from the amplitude and the time delay
of the return signal, respectively, while the target velocity is
measured from the Doppler frequency shift of the received signal
\cite{RichardCurry}. The use of antenna arrays enables improving
the signal strength resulting in improving the accuracy of target
parameter estimation. In particular, antenna arrays are used at
the transmitter to form/steer a beam towards a certain direction
in space yielding coherent processing gain and at the receiver to
coherently process the received data. A corresponding radar system
is commonly referred to as phased-array radar \cite{Skolnik}.
However, it is well-known that phased-array radar suffers from RCS
scintillations which are responsible for signal fading
\cite{Fishler_Haimovich_2}.  Ther\-ef\-ore, a multiple-input
multiple-output (MIMO) radar has recently become the focus of
intensive research \cite{An_idea}--\cite{NasrSergTxEnergyFocusing}.

The essence of the MIMO radar concept is to employ multiple
antennas for emitting several orthogonal waveforms and multiple
antennas for receiving the echoes reflected by the target. MIMO
radar can be either equipped with widely separated antennas
\cite{Radar_separated_review} or colocated antennas
\cite{Radar_colocated_review}. The latter type employs arrays of
closely spaced transmit/receive antennas which results in
increasing the virtual aperture of the receive array due to the
fact that multiple independent waveforms are received by the same
receive array. This enables improving angular resolution,
increasing the upper limit on the number of detectable targets,
and improving parameter identifiability at the price of losing the
transmit coherent processing gain offered by the phased-array
radar \cite{Daum}. On the other hand, a MIMO radar systems with
widely separated transmit/receive antennas enable capturing the
spatial diversity of the target's  RCS
\cite{Radar_separated_review}. Capitalizing on the spatial
diversity of the target, MIMO radar offers a potential to prevent
RCS scintillation and to combat signal fading.

Several techniques are reported in the literature for target
detection and localization in coherent MIMO radar systems
\cite{Fishler_Haimovich_2}, \cite{Bekkerman},
\cite{Xu_Li_Stoica_2008}. However, the main focus of these
techniques is to estimate the directions-of-arrival of targets
located within a certain range-Doppler bin. The problem of
estimating the location and/or velocity of a moving target is
investigated in \cite{Blum_CRB}, \cite{Blum_noncoherent_J}.
However, in some practical applications the target may have
variable speed which necessitates estimating not only the velocity
but also the acceleration of the target. In this case, the target
motion should be modeled as a second order polynomial. In other
cases, even higher order polynomials for modeling the target
motion have to be considered. For example, acceleration and jerk
(rate of acceleration) are used to model the motion of agile
maneuvering targets as described in \cite{JerkModel},
\cite{JerkModel1}. Note that for a maneuvering target, the radial
velocity with respect to any single receiver exhibits variation
which causes significant spread of the radar echo in the Doppler
spectrum \cite{ManeuveringTargets}. Other practical examples that
exhibit variations in the target speed and, therefore, require
high-order target motion modeling, include motion of highly
maneuvering tactical ballistic missiles
\cite{HighlyManeuveringMissile}, landing of fighter jets on
warship carriers \cite{LandingDecisionSupport}, etc.
Unfortunately, the variation in the target's speed in the
aforementioned applications limits the applicability of
conventional techniques for target localization and parameter
estimation.

In this work\footnote{The initial results have been reported in
\cite{SAM10}.}, we develop a maximum likelihood (ML) based
estimator for estimating the parameters of a moving target in
multi-static non-coherent MIMO radar systems. The radar motion
within a certain processing interval is modeled as a general-order
polynomial. In the specific case when the polynomial order is two,
the polynomial coefficients correspond to the initial location,
velocity, and acceleration of the target. By concentrating the ML
function with respect to the nuisance parameters, e.g., reflection
coefficients, we show that the ML problem can be interpreted in
terms of the classic ``overdetermined'' nonlinear least-squares
(LS) problem. The proposed ML estimator requires multi-dimensional
search over the unknown parameters of interest, i.e., the unknown
polynomial coefficients of the target motion. Simulation results
demonstrate an excellent performance of the proposed estimator. It
is worth noting that the superior performance of the proposed
algorithm comes at price of the higher computational complexity
mandated by the ML algorithm. Therefore, the development of a
computationally efficient algorithms that enable reducing the
computational cost of solving the proposed parameter estimation
problem is of interest.

The rest of the paper is organized as follows. The MIMO radar
signal model is given in Section~II while the proposed moving
target motion model is given in Section~III. We derive the ML
estimator in Section~IV. The Cram\'er-Rao Bound (CRB) is derived
in Section~V. Simulation results which show the effectiveness of
the proposed ML estimator are reported in Section~VI followed by
conclusions drawn in Section~VII and Appendix where the details of
CRB derivations are presented. This paper is reproducible research
\cite{Rep} and the software needed to generate the simulation
results will be provided to the IEEE Xplore together with the
paper upon its acceptance.

\section{MIMO Radar Signal Model}
Consider a non-coherent MIMO radar system equipped with $M$
transmit and $N$ receive widely separated antennas. In a Cartesian
three-dimensional (3D) space, the transmit and receive antennas
are assumed to be located at ${\bf p}_m\triangleq [x_m\ y_m\
z_m]^T,\ m=1,\ldots,M$ and ${\bf q}_n\triangleq [x_n\ y_n\
z_n]^T,\ n=1,\ldots,N$, respectively, where $(\cdot)^T$ stands for
the transpose operator. The complex envelope of the signal
transmitted by the $m$-th transmitter can be written as
\begin{equation}\label{eq:mth_Tx_signal}
s_m(t)= \sqrt{\frac{E}{M}}\varphi_m(t),\quad 0\leq t\leq T
\end{equation}
where $E$ is the total transmitted energy, $T$ is the radar pulse
duration, $t$ is the time index within the radar pulse and
$\varphi_m(t)$ is a unit-energy baseband waveform. Waveforms used
at different transmitters are assumed to satisfy the orthogonality
condition \cite{Blum_CRB}
\begin{equation}\label{eq:orthogonal_waveforms}
\int_{T}{\varphi}_m(t){\varphi_i}^{*}(t-\tau)dt = 0,\quad {\rm
for}\ m\neq i,\ \forall \tau
\end{equation}
where $(\cdot)^{*}$ stands for the conjugate operator and $\tau$
is some time delay. Let ${\varphi}_m(t-\tau_m)e^{-j2\pi f_m t}$ be
a time-delayed frequency shifted version of ${\varphi}_m(t)$.
Define the two-dimensional (2D) function $\chi_{m,i}(\tau,\nu)$ as
\begin{eqnarray}\label{eq:chi}
\chi_{m,i}(\tau,\nu)&\!\!\!\triangleq\!\!\!&\!\int_{T}\!{\varphi}_m
\left(t\!-\!
\tau_{m}\right)e^{-j2\pi f_m t}\nonumber\\
&&\cdot{\varphi}_i^{*}(t\!-\!\tau)e^{j2\pi \nu t}dt.
\end{eqnarray}
where $\tau$ and $\nu$ are the time delay and frequency indexes,
respectively. An important property of $\chi_{m,i}(\tau,\nu)$ is
that
\begin{equation}\label{eq:maximum_chi}
\max_{\tau,\nu}|\chi_{m,m}(\tau,\nu)|=1,\quad \tau=\tau_m,\ \nu=f_m.
\end{equation}

\section{Proposed Target Motion Model}
Consider a moving target whose location during the $k$-th radar
pulse is given in the 3D space by
\begin{equation}\label{eq:target_position}
   {\bf L}(k)=[x(k)\ y(k) \ z(k)]^T,\quad k=1,\ldots,K
\end{equation}
where $k$ is the slow time index (i.e. pulse number), and $K$ is
the total number of radar pulses within a certain processing
interval. In (\ref{eq:target_position}), $x(k)$, $y(k)$, and
$z(k)$ are the $x$-, $y-$, and $z$-components of the target
location, respectively. The target location during the $k$-th
radar pulse can be described by the following $Q$-th order
polynomials
\begin{equation}\label{eq:x_position}
    x(k) = \sum_{q=0}^{Q} {C}_q \frac{(kT)^q}{q!} ,\quad k=0,\ldots,K
\end{equation}
\begin{equation}\label{eq:y_position}
    y(k) = \sum_{q=0}^{Q} {D}_q \frac{(kT)^q}{q!} ,\quad k=0,\ldots,K
\end{equation}
\begin{equation}\label{eq:z_position}
    z(k) = \sum_{q=0}^{Q} {E}_q \frac{(kT)^q}{q!} ,\quad k=0,\ldots,K
\end{equation}
where $C_q$, $D_q$, and $E_q$ ($q=0,\ldots,Q$) are the unknown
target motion coefficients and $\cdot!$ stands for the factorial
of an integer. It is worth noting that for the mono-static radar,
the use of~(\ref{eq:x_position})--(\ref{eq:z_position}) to model
the target location results in a polynomial-phase signal (PPS) at
the receiver and leads to the problem of PPS parameter estimation
that has been extensively studied in the literature
\cite{PHAF}--\cite{Gini2}. Note that the order of the PPS can be
higher than two in the case when the carrier frequency at the
transmitter is not constant (e.g., in the case when linear FM
signals are used).

The complex envelope of the signal received by the $n$-th receiver
can be written as
\begin{eqnarray}\label{eq:data_snapshot}
{r}_n(t, k)&\!\!\!\triangleq\!\!\!& e^{j\phi_n}\sum_{m=1}^{M}
\beta_{mn}{s_m}
\left(t-\tau_{mn}(k)\right)e^{j2\pi f_{mn}(k)t}\nonumber\\
&\!\!\! \!\!\!& \cdot\ e^{-j2\pi f_c\tau_{mn}(k)} \!+\! {w}(t,
k), \; k=1,\!\ldots\!,K
\end{eqnarray}
where $\phi_n$ is the unknown initial phase of the $n$-th
receiver, $\beta_{mn}$ is the target reflection
coefficient\footnote{We assume that the reflection coefficient
obeys the Swerling I model, i.e., it remains constant within the
observation interval.} associated with the $mn$-th
transmit-receive path, $f_{mn}(k)$ is the Doppler frequency
associated with the $mn$-th path during the $k$-th pulse, $f_c$ is
the carrier frequency, ${w}(t, k)$ is the independent sensor noise
which is assumed to be zero-mean white circularly Gaussian
process, and $\tau_{mn}(k)$ is the time delay required for the
carrier wave to travel through the $mn$-th transmit-receive path
during the $k$-th pulse. We assume that the signal echoed from the
target is present in the background of clutter plus noise.
Moreover, we assume that the target can only migrate to an
adjacent range-Doppler cell, and, therefore, the characteristics
of the noise plus clutter component remains the same. If the
clutter component is not Gaussian, space-time adaptive processing
(STAP) can be used as a preprocessing step to filter out the
clutter component \cite{KlemmBook}. Note that when the relative
speed between the target and the radar platform is large, the fact
that the spectrum of the clutter is centered around the platform
velocity \cite[Ch.~8]{PrinciplesOfModernRadar} enables the use of
STAP techniques to filter out the clutter.\footnote{In this paper
we assume that the locations of all transmit-receive antennas are
fixed.} The time delay associated with the $(mn)$-th
transmit-receive path can be defined as
\begin{equation}\label{eq_time_delay}
\tau_{mn}(k)\triangleq \frac{\|{\bf L}(k)-{\bf p}_m\|+
\|{\bf L}(k)-{\bf q}_n\|}{c}
\end{equation}
where $c$ is the speed of light.

Using (\ref{eq:orthogonal_waveforms}) and (\ref{eq:maximum_chi}),
the received signal (\ref{eq:data_snapshot}) can be decomposed by
matched-filtering\footnote{In pulsed radar, this process is
commonly referred to as pulse compression. In MIMO radar, it
additionally enables to separate the mixed data at each receive
antenna into components associated with different transmit-receive
paths.} the signal ${r}_n(t, k)$ to the waveforms
${\varphi}_m\left(t\!-\!\tau_{mn}(k) \right) e^{-j2\pi f_{mn} (k)t
}$, yiel\-ding
\begin{eqnarray}\label{eq:matched_filters}
{r}_{mn}(k)
&\!\!\!\!=\!\!\!\!&\!\sqrt{\frac{E}{M}}{\tilde\beta}_{mn}e^{-j2\pi
f_c\tau_{mn}(k)} + w_{mn}(k)
\end{eqnarray}
where ${\tilde\beta}_{mn}\triangleq{\beta}_{mn}e^{j\phi_n}$ and
$w_{mn}(k)$ is the noise component at the output of the matched
filter. Note that the unknown initial phase component is absorbed
in the unknown reflection coefficient. It is worth noting that it
is assumed in \eqref{eq:matched_filters} that range-Doppler cell
synchronization \cite{Maio_Lops} is performed before applying the
matched-filtering step. More specifically, for each radar pulse,
it is assumed that the range-Doppler cell that contains the target
is known. We also assume that the time delay and the Doppler shift
at which the matched-filter is performed coincide with the
location of the peak of (\ref{eq:maximum_chi}). In practice, the
synchronized range-Doppler cell may slightly deviate from the
location of the peak of (\ref{eq:maximum_chi}). To account for the
effect of such a deviation, the  ambiguity function of the
considered MIMO radar should be also considered
\cite{MIMOradarAmbiguity}.

The $MN\times 1$ virtual data vector can be formed as
\begin{eqnarray}\label{eq:virtual_data}
{\bf r}(k) &\!\!\!\triangleq\!\!\!& [{r}_{11}(k),\ldots,{r}_{MN}
(k)]^T \nonumber\\
&\!\!\!=\!\!\!& \sqrt{\frac{E}{M}}{\bf T}(k){\bf b} +
{\bf w}(k)
\end{eqnarray}
where ${\bf T}(k)$ is a $MN\times MN$ diagonal matrix whose
$mn$-th diagonal element is given by $e^{-j2\pi f_c\tau_{mn}(k)}$,
$ {\bf b}\triangleq [{\tilde\beta}_{11},\ldots,$
${\tilde\beta}_{MN}]^T$ is $MN\times 1$ the vector of reflection
coefficients, and $ {\bf w}\triangleq [w_{11}(k),$ $\ldots,
w_{MN}(k)]^T$ is the $MN\times 1$ virtual additive noise term.
Note that each element of ${\bf w}(k)$ has the same statistics as
$w_{mn}(k)$.

\section{Maximum Likelihood Estimation}\label{sec:ML_estimator}
Let the $3(Q+1)\times 1$ vector of unknown coefficients associated
with the moving target be defined as ${\boldsymbol\psi} =
[C_{0},\ldots,\ C_{Q},$ $D_{0}, \ldots,\ D_{Q},$ $E_{0}, \ldots,\
E_{Q}]^T$. Assuming that the reflection coefficients associated
with different transmit-receive paths are constant (deterministic)
values, the virtual observations (\ref{eq:virtual_data}) satisfy
the following statistical model:
\begin{equation}
{\bf r}(k)\sim {\cal N}_C\left(\sqrt{\frac{E}{M}}{\bf T}(k)
{\bf b} , \sigma^2 {\boldsymbol I}\right)
\label{statmod}
\end{equation}
where ${\cal N}_C$ denotes the complex multivariate circularly
symmetric Gaussian probability density function, $\sigma^2$ is
the noise variance, and ${\bf I}$ is the identity matrix.

Then, the negative log-likelihood (LL) function of the unknown
parameters is given as
\begin{eqnarray}\label{eq:netagive_log}
{\mathcal L}({\boldsymbol\psi}, {\bf b}) &\!\!\!=\!\!\!&
\sum_{k=1}^{K} \left\| {\bf r}(k) - \sqrt{\frac{E}{M}} {\bf T}
(k) {\bf b}\right\|^2 \nonumber\\
&\!\!\!=\!\!\!& \sum_{k=1}^{K}{\bf r}^H(k){\bf r}(k) - \left(\sqrt
{\frac{E}{M}}  \sum_{k=1}^{K}{\bf r}^H(k) {\bf T}(k)\right)
{\bf b} \nonumber \\
& & - {\bf b}^H \left(\sqrt{\frac{E}{M}}\sum_{k=1}^{K}{\bf
T}^{H}(k) {\bf r}(k)\right) + {\bf b}^H \left( \frac{E}{M}
\sum_{k=1}^{K} {\bf T}^{H} (k) {\bf T}(k)\right){\bf b}.
\end{eqnarray}
The minimization of \eqref{eq:netagive_log} over ${\bf b}$ yields
\begin{eqnarray}\label{eq:coeff_estimates}
\hat{\bf b} &\!\!\!=\!\!\!& \sqrt{\frac{M}{E}}\left(\sum_{k=1}^{K}
{\bf T}^{H}(k) {\bf T}(k) \right)^{-1} \cdot \left(\sum_{k=1}^{K}
{\bf T}^{H}(k){\bf r}(k)\right)\nonumber\\
&\!\!\!=\!\!\!& \frac{1}{K} \sqrt{\frac{M}{E}} \sum_{k=1}^{K}
{\bf T}^{H}(k){\bf r}(k)
\end{eqnarray}
where the second equality follows from the fact that $(
\sum_{k=1}^{K} {\bf T}^{H}(k) {\bf T}(k) )^{-1} = 1/K{\bf
I}_{MN}$. It is worth noting that \eqref{eq:coeff_estimates} can
be used to compute the RCSs associated with different
transmit-receive paths. This can be employed for reducing the
dimensionality of the data by discarding the data associated with
weak RCSs especially in the case of large values of $M$ and $N$.
Substituting (\ref{eq:coeff_estimates}) into
(\ref{eq:netagive_log}), we obtain
\begin{eqnarray}\label{eq:netagive_log3}
{\mathcal L}({\boldsymbol\psi})&\!\!\!=\!\!\!&\sum_{k=1}^{K}
{\bf r}^H(k) {\bf r}(k) \nonumber\\
& & - \frac{1}{K} \left(\sum_{k=1}^{K}{\bf r}^H(k) {\bf T}(k)\right)
\cdot \left(\sum_{k=1}^{K}{\bf T}^{H}(k){\bf r}(k)\right).
\end{eqnarray}
The target parameters can be estimated by minimizing
\eqref{eq:netagive_log3} over the unknown parameters.
Alternatively, they can be obtained by maximizing the second term
in \eqref{eq:netagive_log3}. Therefore, the ML estimator can be
defined as
\begin{equation}\label{eq:ML_estimator}
\hat{\boldsymbol \psi} = {\rm arg} \min_{\boldsymbol \psi}
{\mathcal L}({\boldsymbol\psi}) = {\rm arg} \max_{\boldsymbol \psi}
\left\|\sum_{k=1}^{K}{\bf T}^H(k){\bf r}(k)\right\|^2.
\end{equation}
The above estimator jointly estimates the target parameters and
generally requires a highly nonlinear optimization of
(\ref{eq:ML_estimator}) over ${\boldsymbol \psi}$. However, if
properly initialized, the optimization of the LL function may be
implemented by means of simple local optimization algorithms.

It is worth noting that the ML estimator can be recast in the form
of the classic ``overdetermined'' nonlinear LS problem. Denoting
$\tilde{\bf r}=[{\bf r}^T (1),\ldots, {\bf r}^T (K)]^T$ and ${\bf
Q}=[{\bf T}^T(1),\ldots, {\bf T}^T(K)]^T$, we can rewrite
(\ref{eq:netagive_log}) as
\begin{equation}\label{eq:netagive_log1}
{\cal L}({\boldsymbol \psi}, {\bf b})=\left\| \tilde{\bf r} -
\sqrt{\frac{E}{M}}{\bf Q} {\bf b}
\right\|^2.
\end{equation}
Minimizing \eqref{eq:netagive_log1} over ${\bf b}$ and
substituting the result in (\ref{eq:netagive_log1}), we obtain
\begin{equation}\label{eq:netagive_log2}
{\cal L}({\boldsymbol \psi}) = \left\|{\bf P}^{\perp}_{{\bf
Q}}{\bf y}\right\|^2=\tilde{\bf r}^H {\bf P}^{\perp}_{{\bf Q}}
\tilde{\bf r}
\end{equation}
where ${\bf P}^{\perp}_{{\bf Q}}={\bf I} - {\bf Q} ({\bf Q}^H{\bf
Q})^{-1}{\bf Q}^H$ is the orthogonal projection matrix onto the
column subspace of ${\bf Q}$. Therefore, the ML estimator can be
re-defined as
\begin{equation}\label{eq:ML_estimator1}
\hat{\boldsymbol \psi} = {\rm arg} \min_{\boldsymbol \psi}
\tilde{\bf r}^H {\bf P}^{\perp}_{{\bf Q}} \tilde{\bf r} = {\rm arg}
\max_{\boldsymbol \psi}
\tilde{\bf r}^H {\bf Q} ({\bf Q}^H{\bf
Q})^{-1}{\bf Q}^H \tilde{\bf r}.
\end{equation}
Note that \eqref{eq:ML_estimator} and \eqref{eq:ML_estimator1} are
equivalent. However, using \eqref{eq:ML_estimator} when optimizing
the ML estimator is computationally more attractive than using
\eqref{eq:ML_estimator1} as it avoids computing the inverse of the
$MN \times MN$ matrix ${\bf Q}^H{\bf Q}$.

Finding the ML estimation based on \eqref{eq:ML_estimator} is in
general difficult and computationally demanding problem especially
for large values of the polynomial order $Q$. Therefore, nonlinear
optimization tools such as genetic algorithms, simulated annealing
based methods, or expectation-maximization (EM)-type procedures
can be used. However, good initialization of such algorithms is
desirable to reduce the complexity. Here, we suggest a simple way
for such an initialization. Particularly, we assume for
initialization that each receive antenna can be used to obtain a
coarse estimate of the target range at the discrete time instants
$k=1,\ldots,K$. Then the coarse estimates to the target range with
respect to different receive antennas can be used jointly to
obtain a coarse estimate to the instantaneous target location
$\hat{\bf L}(k),\ k=1,\ldots,K$. The range-only based target
tracking approach reported in \cite{RangeOnlyTargetTracking} can
be, for example, used. Once, this coarse estimate is obtained, a
simple polynomial regression can be performed to obtained the
polynomial coefficients of the model. The so obtained estimates of
the polynomial coefficients of the target model are then used as
initial values for a specific optimization algorithm used.

\section{Cram\'er-Rao Bound}
In this section, we give explicit expressions for the exact CRB
on the accuracy of estimating the target model parameters. The
$(2MN + 3(Q + 1)) \times 1$ vector of unknown parameters
(including reflection coefficients) can be defined as
\begin{eqnarray}\label{eq:unkown parameters}
{\boldsymbol\Psi} &\!\!\!=\!\!\!& [{\boldsymbol\psi}_x^T,
{\boldsymbol\psi}_y^T, {\boldsymbol\psi}_z^T, {\rm Re}\{{\bf
b}\}^T, {\rm Im}\{{\bf b}\}^T,
\sigma^2]^T \nonumber \\
&\!\!\!=\!\!\!& [{\boldsymbol\psi}^T, \breve{\bf b}^T,
\sigma^2]^T
\end{eqnarray}
where ${\boldsymbol\psi}_x\triangleq [C_0,\ldots,\ C_Q]^T$,
${\boldsymbol\psi}_y\triangleq [D_0,\ldots,\ D_Q]^T$,
${\boldsymbol\psi}_z\triangleq [E_0,\ldots,\ E_Q]^T$,
${\boldsymbol\psi} = [{\boldsymbol\psi}_x^T,
{\boldsymbol\psi}_y^T, {\boldsymbol\psi}_z^T]^T$, and $\breve{\bf
b}=[{\rm Re}\{{\bf b}\}^T, {\rm Im}\{{\bf b}\}^T]^T$.

The elements of the FIM  has the form of the complex circularly
Gaussian process \eqref{statmod} can be expressed as
\begin{equation}\label{eq:FIM1}
{\bf F} = \frac{2E}{\sigma^2 M} \cdot
          \begin{bmatrix}
            {\bf F}_{{\boldsymbol\psi}{\boldsymbol\psi}} &
            {\bf F}_{{\boldsymbol\psi}{\bf b}} \\
            {\bf F}_{{\boldsymbol\psi}{\bf b}}^H & {\bf F}_{\breve{\bf b}
            \breve{\bf b}}  \\
          \end{bmatrix}
\end{equation}
where the $3(Q+1) \times 3(Q+1)$ matrix ${\bf
F}_{{\boldsymbol\psi}{\boldsymbol\psi}}$, the $3(Q+1)\times 2NM$
matrix ${\bf F}_{{\boldsymbol\psi}{\bf b}}$, and the $2NM\times
2NM$ matrix ${\bf F}_{\breve{\bf b}\breve{\bf b}}$ are defined as
follows
\begin{equation}\label{eq:psi_psi}
{\bf F}_{{\boldsymbol\psi}{\boldsymbol\psi}} =
{\rm Re} \left\{ \sum_{k=0}^{K-1}\tilde{\bf B}^H(k)\tilde{\bf T}(k)^H
\tilde{\bf T}(k) \tilde{\bf B}(k) \right\}
\end{equation}
\begin{equation}\label{eq:psi_b}
{\bf F}_{{\boldsymbol\psi}{\bf b}} = {\rm Re}
\left\{ \sum_{k=0}^{K-1} \tilde{\bf B}^H(k) \tilde{\bf T}(k)^H {\bf T}
(k){\boldsymbol J} \right\}
\end{equation}
\begin{equation}\label{eq:b_b}
{\bf F}_{\breve{\bf b}\breve{\bf b}} = {\rm Re} \left\{
\sum_{k=0}^{K-1} {\boldsymbol J}^H {\bf T}^H (k) {\bf T}(k)
{\boldsymbol J} \right\}.
\end{equation}
Derivation of \eqref{eq:psi_psi}--\eqref{eq:b_b} and definitions
of $\tilde{\bf B}^H(k)$, $\tilde{\bf T}(k)$, and ${\boldsymbol J}$
are given in Appendix.

Form (\ref{eq:FIM1}), it follows that the CRB can be obtained as
\begin{eqnarray}\label{eqCRB}
{\bf{CRB}} & \triangleq & \frac{\sigma^2}{2 E/M } \cdot
{\boldsymbol F}^{-1}.
\end{eqnarray}
From \eqref{eqCRB}, we observe that the CRB on estimation
performance is linearly proportional to the noise power and
inversely proportional to the transmitted power per antenna, i.e.,
the CRB is directly proportional to the signal-to-noise ratio
(SNR).

\section{Simulation Results}
In the first example, we assume that there are $M=3$ transmit
antennas in a 2D plane located at $[(0, -5000),$\  $ \ (0, 5000),$
$(5000, 5000)]$m and there are $N=5$ receive antennas located at
$[(0, -5000),$ $(0, 0),$ $(0, 5000),$ $ (2500, 5000),$ $(5000,
5000)]$m. The motion of the target is parameterized by a
second-order motion equation, i.e., by the initial location
$(9800,$ $0)$m, velocity (100,\ 0)m/s, and acceleration $(-20,$
$0)$m/s$^2$. The radar pulse repetition time (PRT) used is
$1.25$ms. The baseband (orthogonal) waveforms used at the three
transmit antennas are exponential harmonics of the frequencies
$1$~KHz, $2$~KHz, and $3$~KHz, respectively. The carrier frequency
$f_c = 300$~MHz is used at all transmit antennas and the
propagation speed is assumed to be $3\times 10^8$~m/s. The
transmitted energy $E$ is normalized so that $\sqrt{E/M}=1$. The
$MN\times 1$ reflection coefficient vector is drawn randomly and
then kept fixed throughout the simulations. The additive noise is
modeled as a complex Gaussian zero-mean unit-variance spatially
and temporally white process that has identical variances in each
receive antenna. The whole observation time used in $0.5$s and is
assumed to be divided into $Z=50,\ (Z < K)$ equally spaced
intervals of width $0.01$s each, where $K$ was introduced earlier
to denote the number of radar pulses. Each interval is assumed to
be a coherent integration time (CIT), i.e., every CIT contains
eight radar pulses. It is observed that the difference between the
Doppler frequencies associated with the first and the last radar
pulses within a certain CIT does not exceed $0.0013$~Hz for all
CITs within the whole observation time. Therefore, it can be
assumed that the Doppler frequency does not change during the same
CIT but changes from CIT to CIT.\footnote{For scenarios that
involve rapid change in the target speed such as a highly
maneuvering target, the duration of the CIT should be reduced. The
shortest CIT duration that can be used is one PRT. However, this
comes at the price of higher number data samples, i.e., the number
of intervals $Z$. This leads to a higher computational cost.} The
ML estimator (\ref{eq:ML_estimator}) is used to estimate the
target parameters. Instead of finding the minimum of
(\ref{eq:netagive_log2}), we search for the peak of the positive
LL function
\begin{equation}
{\cal L}_p({\boldsymbol \psi}) = {\bf y}^H{\bf P}_{{\bf Q}}{\bf y}
\end{equation}
where ${\bf P}_{{\bf Q}} \triangleq {\bf Q} ({\bf Q}^H{\bf
Q})^{-1}{\bf Q}^H$. The  genetic algorithm (GA) is used to
optimize ${\cal L}_p({\boldsymbol \psi})$ over the unknown
parameters $C_q$, $q=0,\ 1,\ 2$, i.e., the unknown target initial
location, velocity, and acceleration. To make sure that the
estimation accuracy is not limited by the size of the search
region, the boundaries of the GA search region, for each
parameter, are taken wide enough ($20$ times larger than the
corresponding CRB) and centered at the true values. The root
mean-square errors (RMSEs) are computed for the parameters of
interest based on $100$ independent simulation runs. The RMSEs of
the estimates of the unknown parameters are compared to the
corresponding CRBs.

\begin{figure}[h!]
\centering \centerline{\epsfig{figure=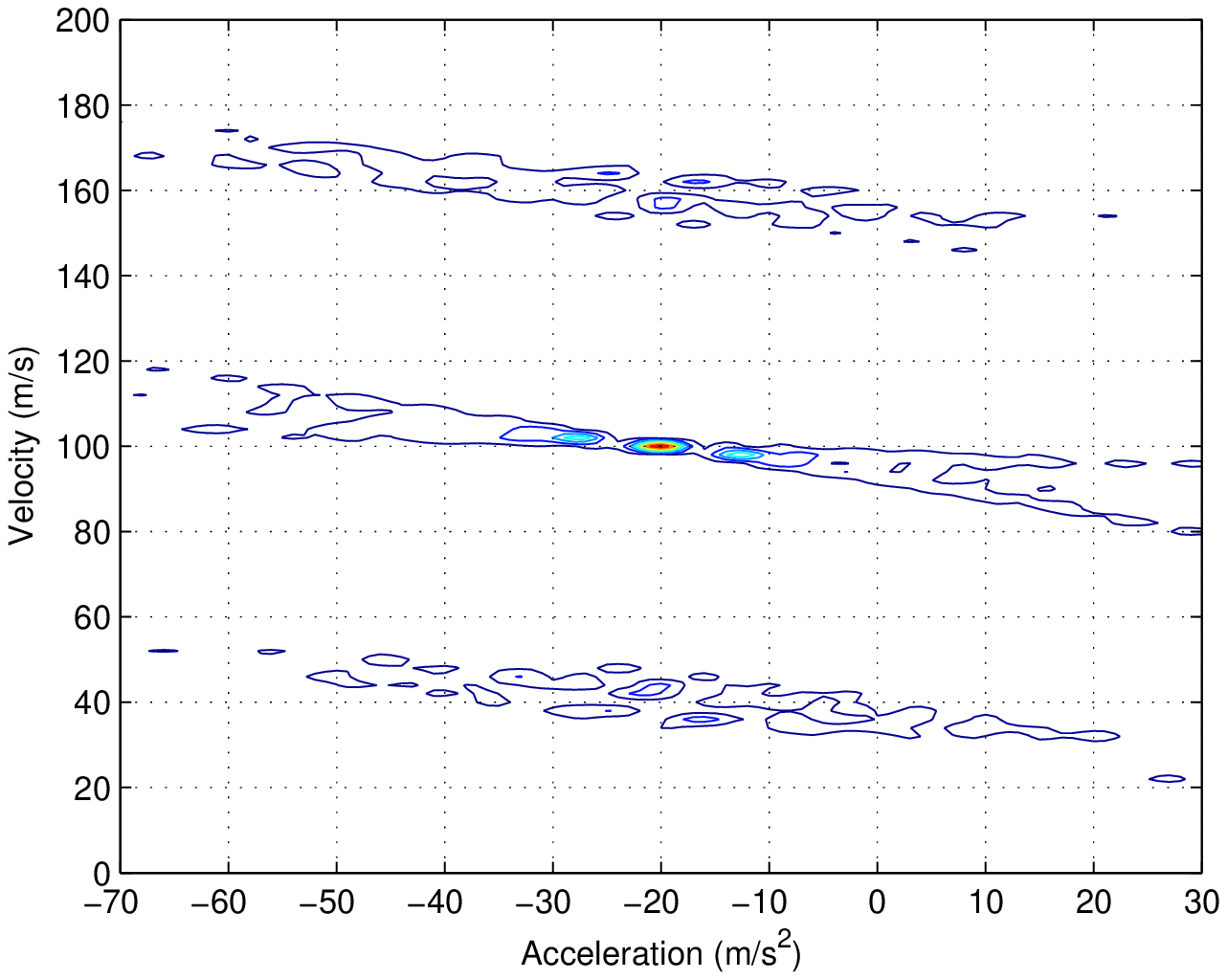,width=11cm}}
\caption{Contour plot of the ML function \eqref{eq:ML_estimator1}
in the velocity-acceleration plane; example~1.}
\label{fi:contour}
\end{figure}

Fig.~\ref{fi:contour} shows the contour plot of
(\ref{eq:ML_estimator}) computed in the 2D velocity-acceleration
plane while the initial location is fixed to its true value. The
SNR for this case is fixed to $0$~dB. It can be seen from this
figure that the ML estimator exhibits main peak close to the true
values of both the velocity and acceleration parameters. Two other
2D contour plots computed in the location-velocity and
location-acceleration planes exhibit similar behavior as that in
Fig.~\ref{fi:contour}. The location-velocity and
location-acceleration contour plots are similar.

\begin{figure}[h!]
\centering \centerline{\epsfig{figure=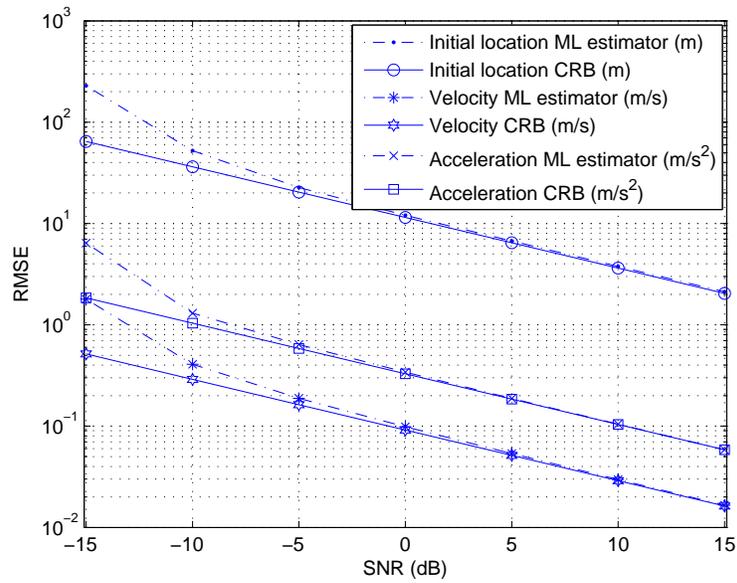,width=11cm}}
\caption{RMSEs versus SNR; example~1.}
\label{fi:RMSE}
\end{figure}

Fig.~\ref{fi:RMSE} shows the RMSEs versus SNR for the initial
location, velocity and acceleration. It can be seen from the
figure that the initial location\footnote{Note that the initial
location corresponds to the location during the $1$-st pulse,
i.e., at $k=0$. The location at $k$-th time instant within the
observation interval can be easily computed by substituting the
estimated values of the polynomial coefficients corresponding to
initial location, velocity, and acceleration in
\eqref{eq:x_position}--\eqref{eq:z_position}.} estimation accuracy
is in the range of tens of meters at SNR values below $0$~dB and
it is in the range of meters at SNR values above $0$~dB. Also, it
can be observed from the figure that the RMSEs for the initial
location, velocity, and acceleration estimation coincide with the
CRB at moderate and high SNR regions. It is clear from
Fig.~\ref{fi:RMSE} that the proposed ML estimator offers excellent
estimation accuracy for estimating the target location, velocity,
and acceleration.

In the second example, we show that the proposed method is also
applicable to the case of fixed speed targets. In this case, the
target motion is described by a first-order polynomial where the
initial location of the target is taken as $(8400,$ $9800)$m and
the target velocity is assumed to be (40,\ -50)m/s. The transmit
antennas are located at $[(0, 0),$\  $ \ (4000, 0),$ $(0, 4000)]$m
and the receive antennas are located at $[(0, 0),$ $(2000, 0),$
$(0, 2000),\ (6000, 0),$ $(0, 6000)]$m. The radar pulse width and
the waveforms used at the transmitters are the same as in the
first example.  The overall observation duration is $2.0$s. Noting
that the target speed is constant, the Doppler frequency is the
same during the whole observation time which enables using longer
CITs. The observation time is divided into $K=50$ equally spaced
intervals of duration $0.04$s each. Each CIT involves energy
integration over $32$ radar pulses. Similar to the previous
example, the GA initialized around the true parameters is used to
optimize the LL function over the unknown initial location and
target velocity components.

\begin{figure}[h!]
\centering
\centerline{\epsfig{figure=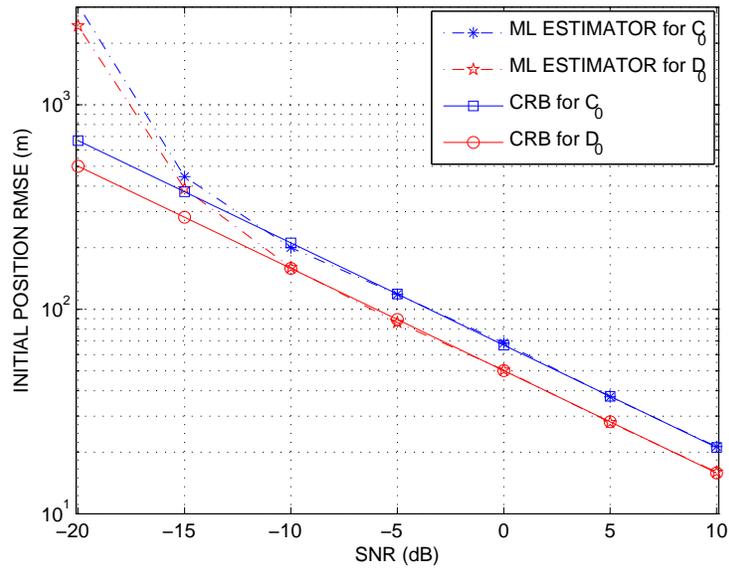,width=11cm}}
\caption{Initial position estimation RMSEs versus SNR; example~2.}
\label{fi:RMSE4Dposition}
\end{figure}

\begin{figure}[h!]
\centering
\centerline{\epsfig{figure=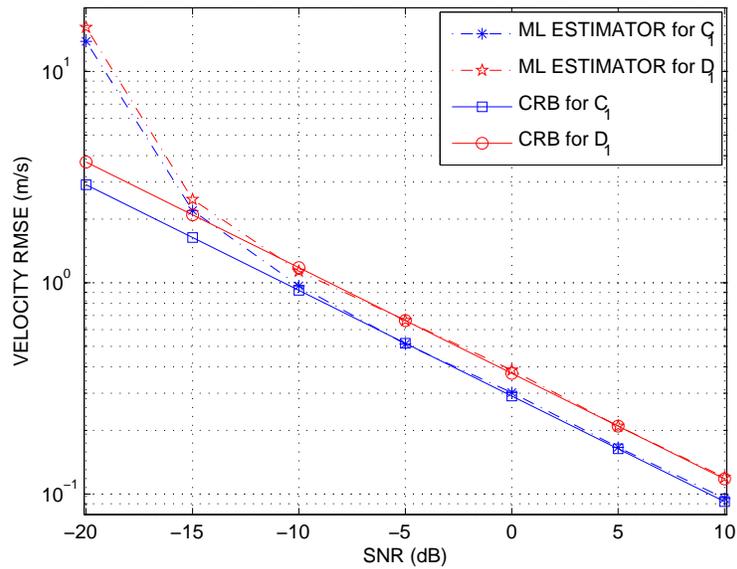,width=11cm}}
\caption{Velocity estimation RMSEs versus SNR; example~2.}
\label{fi:RMSE4dvelocity}
\end{figure}

Fig.~\ref{fi:RMSE4Dposition} shows the RMSEs versus SNR for the
$x$- and the $y$-components of the target initial location. It can
be seen from this figure that the performance of the proposed ML
method coincides with the CRB for SNR values higher than $-10$~dB.
Fig.~\ref{fi:RMSE4dvelocity} shows the RMSEs versus SNR for the
$x$- and the $y$-components of the target velocity. It can be seen
from the figure that the proposed ML method has excellent velocity
estimation performance which coincides with the CRB as the  SNR
increases.

\section{Conclusions}
A new ML estimator for moving target parameter estimation in
non-coherent MIMO radar has been developed. The target motion
within a certain processing interval is modeled as a general-order
polynomial which is suitable for modeling the motion of a moving
target with rapidly changing speed such as a jet landing on an
aircraft carrier. The ML function is concentrated with respect to
the nuisance parameters (target reflection coefficients). The
resulting ML estimator requires a multi-dimensional search over
the unknown parameters of interest (coefficients of the target
motion model). It has been shown that the proposed ML approach can
be interpreted in the form of the classic ``ove\-r\-d\-etermined''
nonlinear LS problem. The performance of the proposed ML estimator
is validated by simulations and it is shown that it achieves the
CRB derived for the considered parameter estimation problem.

\section*{Appendix: Computation of the Fisher Information Matrix}
The elements of the FIM of a complex circularly Gaussian process
${\bf x}(k)\sim {\cal N}_C\left({\boldsymbol \mu}(k), {\boldsymbol
R}\right)$ are given by \cite{StoicaNehorai}
\begin{eqnarray}
\left[{\boldsymbol F}\right]_{i, j} &=& N\, {\rm trace}\left\{
{\boldsymbol R}^{-1} \frac{\partial {\boldsymbol R}}{\partial \Psi_i}
{\boldsymbol R}^{-1} \frac{\partial {\boldsymbol R}}
{\partial \Psi_j} \right\}\nonumber\\
&& + \, 2\, {\rm Re} \left\{ \sum_{k=0}^{K-1} \frac{\partial
{\boldsymbol \mu}^H(k)}{\partial \Psi_i}
{\boldsymbol R}^{-1} \frac{\partial {\boldsymbol \mu}(k)}
{\partial \Psi_j} \right\}
\label{CRB}
\end{eqnarray}
where $\Psi_i$ is the $i$-th element of ${\boldsymbol\Psi}$.
Applying (\ref{CRB}) to the model (\ref{statmod}), we obtain
\begin{eqnarray}
\left[{\boldsymbol F}\right]_{i, j} & = & \frac{K N M}{\sigma^4}\,
\frac{\partial \sigma^2}{\partial \Psi_i}\,
\frac{\partial \sigma^2}{\partial \Psi_j} \nonumber\\
&&  + \, \frac{2E}{\sigma^2 M} {\rm Re}\left\{
\sum_{k=0}^{K-1}
\frac{\partial\left\{{\bf b}^H
{\bf T}^H(k)\right\}}
{\partial \Psi_i}\cdot \,
\frac{\partial\left\{{\bf T}(k) {\bf b}\right\}}{\partial \Psi_j}
\right\}
\label{CRBpart}
\end{eqnarray}

Direct computation yields,
\begin{eqnarray}
 \frac{\partial\left\{{\bf b}^H {\bf T}^H(k)\right\}}{\partial
 {\rm Re}\{{\bf b}\}} &=& {\bf T}^H(k)\label{eq:partial_Re_b} \\
 \frac{\partial\left\{{\bf b}^H {\bf T}^H(k)\right\}}{\partial
 {\rm Im}\{{\bf b}\}} &=& - j {\bf T}^H(k)\label{eq:partial_Im_b} .
\end{eqnarray}
Introducing the $NM\times 2NM$ matrix ${\boldsymbol J} = [
{\boldsymbol I}, j {\boldsymbol I} ]$, we can rewrite
(\ref{eq:partial_Re_b}) and (\ref{eq:partial_Im_b}) in a compact
form as
\begin{eqnarray}
\frac{\partial \left\{ {\bf b}^H{\bf T}^H (k) \right\} }{\partial
\breve{\bf b}}
 =  {\boldsymbol J}^H {\bf T}^H (k).
\label{DerivativeOver_b}
\end{eqnarray}

Eq.~(\ref{eq_time_delay}) can be rewritten as
\begin{equation}\label{eq_time_delay1}
\tau_{mn}(kT)\triangleq \frac{d_m + d_n}{c}
\end{equation}
where
\begin{eqnarray}
  d_m &=& \left[\left(\sum_{q=0}^{Q} {C}_q \frac{(kT)^q}{q!}
  - x_m\right)^2 \right. \nonumber\\
    && \left. + \left(\sum_{q=0}^{Q} {D}_q \frac{(kT)^q}{q!}
    - y_m\right)^2  + \left(\sum_{q=0}^{Q} {E}_q \frac{(kT)^q}{q!}
    - z_m\right)^2\right]^{1/2} \\
  d_n &=& \left[\left(\sum_{q=0}^{Q} {C}_q \frac{(kT)^q}{q!}
  - x_n\right)^2 \right. \nonumber\\
    && \left. + \left(\sum_{q=0}^{Q} {D}_q \frac{(kT)^q}{q!}
    - y_n\right)^2 + \left(\sum_{q=0}^{Q} {E}_q \frac{(kT)^q}{q!}
    - z_n\right)^2\right]^{1/2}.
\end{eqnarray}
Straightforward computations yield
\begin{eqnarray}
\frac{\partial \left\{ e^{-j2\pi f_c\tau_{mn}(kT)} \right\} }{\partial C_0}
&\!\!\!=\!\!\!& \frac{\partial \left\{ e^{-j2\pi f_c\tau_{mn}(kT)} \right\} }
{\partial \tau_{mn}(kT)}\cdot\frac{\partial \tau_{mn}(kT)}{\partial C_0}
\nonumber\\
&\!\!\!=\!\!\!& \left\{-j2\pi f_c e^{-j2\pi f_c\tau_{mn}(kT)}\right\}
\nonumber\\
&\!\!\!\!\!\!& \cdot\frac{1}{c}\left(\frac{\sum_{q=0}^{Q} {C}_q
\frac{(kT)^q}{q!} - x_m}{d_m}  + \frac{\sum_{q=0}^{Q} {C}_q
\frac{(kT)^q}{q!} - x_n}{d_n}\right) \label{derive_x0}
\end{eqnarray}
Therefore, we obtain
\begin{eqnarray}
\frac{\partial \left\{ {\bf b}^H{\bf T}^H (k) \right\} }{\partial C_0}
&\!\!\!=\!\!\!& {\bf b}^H \left({\bf T}(k)\odot{\bf Z}_x(k)\right)^H
\label{derive_x0_1}
\end{eqnarray}
where the $MN\times MN$ diagonal matrix ${\bf Z}_x(k)$ is given by
\begin{eqnarray}\label{eq:Z_x}
[{\bf Z}_x(k)]_{nM+m, nM+m} &\!\!\!=\!\!\!&  \frac{-j2\pi f_c }{c}
\left(\frac{\sum_{q=0}^{Q} {C}_q \frac{(kT)^q}{q!} - x_m}{d_m}
\right. \nonumber\\
&\!\!\!\!\!\!& \left. + \frac{\sum_{q=0}^{Q} {C}_q \frac{(kT)^q}{q!}
- x_n}{d_n}\right)
\end{eqnarray}
Similar computations yield
\begin{eqnarray}
\frac{\partial \left\{ {\bf b}^H{\bf T}^H (k) \right\} }{\partial C_q}
&\!\!\!=\!\!\!& \frac{(kT)^q}{q!}{\bf b}^H \left({\bf T}(k)
\odot{\bf Z}_x(k)\right)^H,\quad q=1,\ldots,Q.
\label{derive_vx}
\end{eqnarray}

Introducing the $(Q+1) \times 1$ vector ${\bf h}\triangleq
[1,\ldots,\ \frac{(kT)^Q}{q!}]^T$, we can define the $2NM\times
(Q+1)$ matrix
\begin{equation}\label{eq:B_k}
{\bf B}(k) = {\bf h}^T\otimes{\bf b}.
\end{equation}
Therefore, we obtain
\begin{eqnarray}
\frac{\partial \left\{ {\bf b}^H{\bf T}^H (k) \right\} }{\partial
{\boldsymbol\psi}_x}
&\!\!\!=\!\!\!& {\bf B}^H(k)\tilde{\bf T}^H_x(k).
\label{derive_psi_x}
\end{eqnarray}
where $\tilde{\bf T}_x(k)\triangleq{\bf T}(k)\odot{\bf Z}_x(k)$.

Following the same steps, we obtain
\begin{eqnarray}
\frac{\partial \left\{ {\bf b}^H{\bf T}^H (k) \right\} }{\partial
{\boldsymbol\psi}_y}
&\!\!\!=\!\!\!& {\bf B}^H(k)\tilde{\bf T}^H_y(k).
\label{derive_psi_y}
\end{eqnarray}
where $\tilde{\bf T}_y(k)\triangleq{\bf T}(k)\odot{\bf Z}_y(k)$
and the $MN\times MN$ diagonal matrix ${\bf Z}_y(k)$ is given by
\begin{eqnarray}\label{eq:Z_y}
[{\bf Z}_y(k)]_{nM+m, nM+m} &\!\!\!=\!\!\!&  \frac{-j2\pi f_c }{c}
\left(\frac{\sum_{q=0}^{Q} {D}_q \frac{(kT)^q}{q!}-y_m}{d_m} \right.
\nonumber\\
&\!\!\!\!\!\!& \left. + \frac{\sum_{q=0}^{Q} {D}_q \frac{(kT)^q}{q!}
- y_n}{d_n}\right).
\end{eqnarray}
Similarly, we have
\begin{eqnarray}
\frac{\partial \left\{ {\bf b}^H{\bf T}^H (k) \right\} }{\partial
{\boldsymbol\psi}_z}
&\!\!\!=\!\!\!& {\bf B}^H(k)\tilde{\bf T}^H_z(k).
\label{derive_psi_y}
\end{eqnarray}
where $\tilde{\bf T}_z(k)\triangleq{\bf T}(k)\odot{\bf Z}_z(k)$
and the $MN\times MN$ diagonal matrix ${\bf Z}_z(k)$ is given by
\begin{eqnarray}\label{eq:Z_y}
[{\bf Z}_z(k)]_{nM+m, nM+m} &\!\!\!=\!\!\!&  \frac{-j2\pi f_c }{c}
\left(\frac{\sum_{q=0}^{Q} {E}_q \frac{(kT)^q}{q!}-z_m}{d_m} \right.
\nonumber\\
&\!\!\!\!\!\!& \left.
 + \frac{\sum_{q=0}^{Q} {E}_q \frac{(kT)^q}{q!} - z_n}{d_n}\right).
\end{eqnarray}

Introducing the matrix ${\tilde{\bf B}(k)}\triangleq{\bf
I}_3\otimes{\bf B}(k)$ and the matrix $\tilde{\bf T}(k) =
[\tilde{\bf T}_x(k),\ \tilde{\bf T}_y(k),\ \tilde{\bf T}_z(k)]$,
we obtain
\begin{eqnarray}
\frac{\partial \left\{ {\bf b}^H{\bf T}^H (k) \right\} }{\partial
{\boldsymbol\psi}}
&\!\!\!=\!\!\!& \tilde{\bf B}^H(k)\tilde{\bf T}^H(k).
\label{DerivativOverPsi}
\end{eqnarray}

Substituting \eqref{DerivativeOver_b} and \eqref{DerivativOverPsi}
in  (\ref{CRBpart}), the expressions that define the FIM in
\eqref{eq:FIM1}--\eqref{eq:b_b} are readily obtained.

\section*{Acknowledgment}
The authors would like to thank Dr.~Michael Ruebsamen from
Darmstadt University of Technology for helpful discussion on ML
optimization using local search techniques.

\newpage

\end{document}